\RequirePackage{rotating}
\documentclass[acmsmall,prologue,svgnames,nonacm]{acmart}

\usepackage[shortlabels]{enumitem}
\usepackage[english]{babel} 
\usepackage{blindtext}

\usepackage{rotating}[figuresleft]

\usepackage{multirow}

\acmJournal{FACMP}
\setcopyright{rightsretained}
\copyrightyear{2024}
\acmYear{2024}

\begin{document}

\title{\emph{TimeFlows}: Visualizing Process Chronologies from Vast Collections of Heterogeneous Information Objects}

\author{Max Lonysa Muller}
\authornote{Corresponding Author}
\email{max.muller@nationaalarchief.nl}
\orcid{0000-0001-5707-6944}
\affiliation{%
	\institution{Nationaal Archief}
	\city{Den Haag}
	\country{The Netherlands}
}

\author{Erik Saaman} 
\email{erik.saaman@nationaalarchief.nl}
\affiliation{%
	\institution{Nationaal Archief}
	\city{Den Haag}
	\country{The Netherlands}
}

\author{Jan Martijn E. M. van der Werf}
\email{j.m.e.m.vanderwerf@uu.nl}
\affiliation{%
	\institution{Utrecht University}
	\city{Utrecht}
	\country{The Netherlands}
}

\author{Charles Jeurgens}
\email{k.j.p.f.m.jeurgens@uva.nl}
\affiliation{%
	\institution{Universiteit van Amsterdam}
	\city{Amsterdam}
	\country{The Netherlands}
}

\author{Hajo A. Reijers}
\email{h.a.reijers@uu.nl}
\affiliation{%
	\institution{Utrecht University}
	\city{Utrecht}
	\country{The Netherlands}
}

\begin{abstract}
In many fact-finding investigations, notably parliamentary inquiries, process chronologies are created to reconstruct how a controversial policy or decision came into existence. Current approaches, like timelines, lack the expressiveness to represent the variety of relations in which historic events may link to the overall chronology. This obfuscates the nature of the interdependence among the events, and the texts from which they are distilled. Based on explorative interviews with expert analysts, we propose an extended, rich set of relationships. We describe how these can be visualized as \textit{TimeFlows}. We provide an example of such a visualization by illustrating the Childcare Benefits Scandal -- an affair that deeply affected Dutch politics in recent years. This work extends the scope of existing process discovery research into the direction of unveiling non-repetitive processes from unstructured information objects.  
\end{abstract}


\maketitle              

\newcommand{\mywidth}[1]{}

\section{Introduction} \label{introduction}
In various situations, it is worthwhile to determine retrospectively how a process has unfolded over time. The established discipline of \textit{process mining} provides algorithms for discovering \textit{work processes} from historic data. Such processes are cyclic in nature and can be discovered on the basis of events generated by IT systems. However, there are also processes worth reconstructing that are non-repetitive and lack a log of structured events. 

Consider a \emph{parliamentary inquiry}, which is an important use case for the work presented in the paper. A parliamentary inquiry is initiated to gain knowledge on a particular topic when a matter of national importance is at stake. The objective is to reconstruct the process that led to severe consequences, either for the government itself or for the citizens involved. The process in question, once reconstructed, is typically represented in the form of a\textit{ process chronology}, 
e.g. a timeline, a textual summary, or a fact sheet of events. The purpose of a process chronology is to capture what relevant events happened at different moments in time and how these events relate to one another.  

 
The generation of a process chronology, either in the context of a parliamentary inquiry or any other sort of fact-finding investigation, may need to take place on the basis of a wide range of information objects. While such objects may be structured, it is often the case that relevant events are included in emails, text messages, policy briefs, reports, and other unstructured information objects. The analysis of such events is time-consuming as it is mostly done by hand and relies heavily on domain knowledge. A major challenge for these analysts is to present the process of interest in a concise and intuitive way. The objective of this paper is threefold: (1) to conceptualize the the events and texts of a 
process chronology
in a meaningful way, (2) to provide an intuitive visualization thereof, and (3) to set an agenda for 
process chronology
research paving the way for automation. We view this work as a first step of a broader ambition to study requirements, design, and the automated generation of interactive process chronologies. 

Our proposal links to earlier work, which deals with the generation of \textit{timelines} on the basis of a set of textual documents as input. The Timeline Summarization (TLS) line of research aims at automatically identifying key dates of major events from a corpus of texts along with short descriptions of what happened on these dates \cite{steen2019}. Although applicable in settings where the datasets are small and homogeneous, a drawback of these linear timelines is that they quickly become difficult to interpret, e.g. when the amount of dates grows too large or the textual topics are heterogeneous \cite{yu2021}. 

\begin{figure}[H]
\centering
\includegraphics[width=0.64\textwidth]{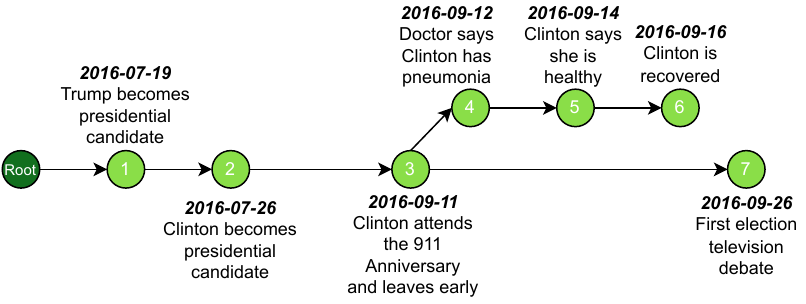}
\caption{This US elections TGR \cite{liu2020} links events through the \emph{Subject Relation}.}
\label{storytree2016}
\end{figure}

More recently, in an attempt to break away  from the simple, linear structure provided by TLS, \emph{timelines with graph-based representations} (TGRs) were proposed. These structures provide a more intricate structure for the representation of related events (e.g. Figure \ref{storytree2016}). Pivotal work in this area was carried out in the early 2010s by Shahaff and others \cite{shahaf2010}. A recent survey on the topic is provided in \cite{noram2023}. A drawback of TGRs, however, is that they center around one type of relation that ties the various events together. In reality, a wide range of relationships, e.g. causality, topicality, time order, etc., may be required to show how events contribute to the overall chronology of the process. In addition, the TGRs fail to relate to the original information objects that they are based on.

In this paper, we propose \emph{TimeFlows} as a way to visually represent process chronologies. They extend TGRs to represent collections of related events. Our proposal improves upon the state of the art in that it (1) provides flexibility for the inclusion of a wide range of relationships between events and (2) includes references to the constitutive information objects for the presented events. TimeFlows are applicable within a wide range of applications, including historical and forensic research. Their objective is to come up with a concise yet intuitive reconstruction of a process based on a heterogeneous and potentially large set of information objects. 

This paper is structured as follows. In Section~\ref{literature}, we delve deeper into related work. Section \ref{researchmethod} outlines our research method. It describes our approach to identify the types of relationships that are important for a process chronology through a set of structured interviews with expert analysts. Then, in Section \ref{timeflows}, we (1) present the results obtained by exploratively coding the interviews, (2) motivate our choice of relationships to be included, and (3) provide a Process Delivery Diagram, which  depicts how manual constructions of process chronologies takes place. In Section \ref{framework}, we present a model for process chronologies and describe how TimeFlows can be used to visualize these. We illustrate our proposal by presenting a TimeFlow for the Dutch Childcare Benefits Scandal~\footnote{See: \url{https://en.wikipedia.org/wiki/Dutch_childcare_benefits_scandal}}. 
We conclude the article with an agenda for further research (Section~\ref{challenges}).

\section{Literature Review} \label{literature}
Different models for TGRs have been proposed \cite{noram2021} \cite{ansah2019}, all of which aim to represent what we have coined \emph{process chronologies}. 
Several algorithms have been devised to automatically generate such graphs, as highlighted in a recent overview~\cite{noram2023}. It includes the approaches described in~\cite{shahaf2010} and \cite{xu2013}. 

The authors of \cite{noram2021} anchor their definition of a \emph{narrative map} (a synonym of a temporal graph) by means of a weighted, directed, acyclic graph. Edges between events are made solely on the basis of topical or thematic coherence. The degree of coherence determines the weight assigned the edges.

In \cite{ansah2019}, the authors define yet another synonym for a temporal graph, namely a \emph{Story Graph}. It is considerably more elaborate and complicated than the aforementioned definition. The authors define notions such as \emph{Related Events} (by means of a user-defined coherence threshold), \emph{Topic Theme}, and \emph{Timeline Path}.  
The authors of \cite{liu2020} graphically depict four different types of structures to characterize a story. These consist of (a) a flat structure, (b) a timeline structure, (c) a graph structure, and (d) a tree structure. In all of these are graphs the nodes indicate events, while the edges denote a topical coherence among these events. 
As noted in \cite{noram2023}, some frameworks allow for forms of interactivity. For instance, in \cite{shahaf2010}, the authors enable the functionality of zoomability of the events, to inspect them at finer or coarser levels of granularity. Moreover, the authors of \cite{shahaffinfomaps} propose the introduction of a \emph{personalized coverage function} that can be employed to display temporal graphs that fit the users' needs and preferences, and suit their own queries. 

What these approaches have in common, is that they all assume there is only one type of edge - a topical or thematic connection between events. Yet recent research has shown that analysts make different types of connections between events \cite{noram2022}. For instance, they also  construct connections on the basis of causality, entities, or specific domain knowledge. In the present paper, we seek to find out which types of relations are most relevant to analysts, and propose a visualization technique process chronologies that captures the variety of relevant relations among events, called \emph{TimeFlows}. Moreover, we investigate how to integrate the documents or texts -- more broadly: information objects -- on which these events are based within our conceptualization as well. 


\section{Research Method} \label{researchmethod}
We aim at grounding our proposal in data obtained through explorative, semi-structured interviews. These interviews are conducted with people having experience with manually constructing process chronologies in their professional work in their own, specific contexts. By analyzing the answers they have provided us, we seek to determine what kind of challenges analysts face when manually constructing process chronologies, in particular with respect to the relations they wish to establish between relevant events. 

We collect data by recording interviews with four interviewee groups. These interviews are transcribed and exploratively coded. Based on the results, we construct a Process Delivery Diagram (PDD) in Section 3 and present our proposal for TimeFlows in Section 4. 
\mywidth{-4mm}
\subsection{Interviewee Groups}
We selected interviewee groups on the basis of their professional experience with creating process chronologies. 
All participants have been actively involved in the creation of process chronologies based on unstructured information objects; three of them had an affinity with parliamentary inquiries.
We conducted four interviews for our data collection: 
\begin{enumerate}[noitemsep]
    \item[(I1)] Four researchers of a Dutch Ministry that have experience in creating chronologies for Parliament.
    \item[(I2)] An interview with a junior council advisor who assisted a minister before and during the hearings for a parliamentary inquiries. 
    \item[(I3)] An investigative journalist who works for a well-known Dutch publication platform.
    \item[(I4)] Two national government officials that have experience as a coordinator of several parliamentary inquiry research staff.
\end{enumerate}


\noindent These interviewee groups are referred to as (I1) through (I4) throughout this paper. When we quote someone from a particular group, it is implied that their views correspond to the consensus opinion of all individuals comprising that particular group. 

\mywidth{-4mm}
\subsection{Semi-Structured Interviews}


The interviews were conducted in Dutch and focused on the approach of the different interviewee groups to create process chronologies. In particular, we asked about the different types of \emph{relationships} the interviewees made between and among information objects and events, in order to be able to create meaningful chronologies.

The questions consisted of the following five sections. We list them below, and describe what kind of questions correspond to each section: 

\begin{enumerate}[noitemsep]
    \item \underline{\emph{Professional Information}}. Interviewees were asked about job title, typical workday activities, and years of experience. 
    \item \underline{\emph{Activities and Process}}. These questions pertained to what activities their occupation consisted of. We also asked them specifically whether their job entailed making chronologies for parliamentary inquiries or other fact-finding investigations. If that was the case, we probed them about these activities, and asked how they conducted them. 
    \item \underline{\emph{Textual Analysis}}. In this section of the interview, we asked participants to describe the different steps they go through to discover and analyse texts they need to read in order to be able to create chronologies or policy document reconstructions. 
    \item \underline{\emph{Relationships between Texts}}. Questions within this section pertained to the relationships professionals make between texts while conducting research.  
    \item \underline{\emph{Comparison between Texts and Events}}. By asking these questions, we aimed at gaining an understanding of how the interviewees interpreted the level of granularity at which events correspond to texts. 
    \item \underline{\emph{Operations}}. This section would delve into the groups' tactics to single out particular passages in texts -- like summarizing, highlighting, or splitting.
\end{enumerate}

\noindent Overall, these questions were aimed at comprehending how the analysts carry out their day-to-day activities. We were specifically interested in what tactics and techniques the interviewees employ to create process chronologies, and what kinds of relations bore the most relevance to their work.

The following relationships were discussed during the interview. A substantial portion of them were based on the work in \cite{noram2022}. Other relationships were also incorporated in the interviews. The ones that form additions to the work of the authors of \cite{noram2022} are marked with an asterisk (\textbf{*}). In this context, A and B refer to distinct events.

\begin{itemize}[noitemsep]
    \item Temporal (A took place before B). 
    \item Similarity (A is similar to B). 
    \item Entity (A is about the same entity -- a person, organization, or place -- as B) 
    \item Citation (A refers to B by means of a hyperlink to another document) 
    \item Correspondence* (person P mentions A in a message to person Q, whom subsequently refers to B in a returned or forwarded message) 
    \item Subject (A falls under the same theme or topic as B) 
    \item Causal (A leads to B) 
    \item Social* (A and B are discussed among persons $P_{1}, \dots, P_{n}$)
    \item Speculative (A is connected to B) 
    \item Domain knowledge (A is related to B through external domain knowledge X, which is not found or represented in the texts themselves) 
    \item Categorical/Ontological* (A is filed under the same category or folder as B, for instance in a DMS or a file structure) 
\end{itemize}

The interviewees were asked to (i) state which relationships they employed during their efforts to manufacture chronologies, and (ii) name the top three of most important relations (in no particular order). Sometimes, they would mention just two important relationships; on other occasions they named four of them. \par

\section{Constructing Process Chronologies}\label{timeflows}
In this section, we present the results of exploratively coding the interviews with the four interviewee groups. 
Based on the interviews, we constructed a Process Delivery Diagram (PDD) that describes how professionals currently construct chronologies. 
A PDD combines a process model and a conceptual model to indicate which concepts are created in which activities~\cite{vandeweerd_2006_PDD}. 
Activities in a PDD can have unordered subactivities to express that there is no predefined sequence in which these need to be carried out.
In Section~\ref{sec:situatianalisedTimeflow}, we describe how process chronologies are currently constructed,
in Section~\ref{sec:foundrelations}, we discuss what relations are used, both implicitly and explicitly, to extract concepts -- such as events and entities -- from the collected information objects.


\mywidth{-4mm}

\subsection{Situationalized Process Chronologies}\label{sec:situatianalisedTimeflow}
\mywidth{-1mm}

\begin{figure}[tb]
\centering
\includegraphics[width=\textwidth]{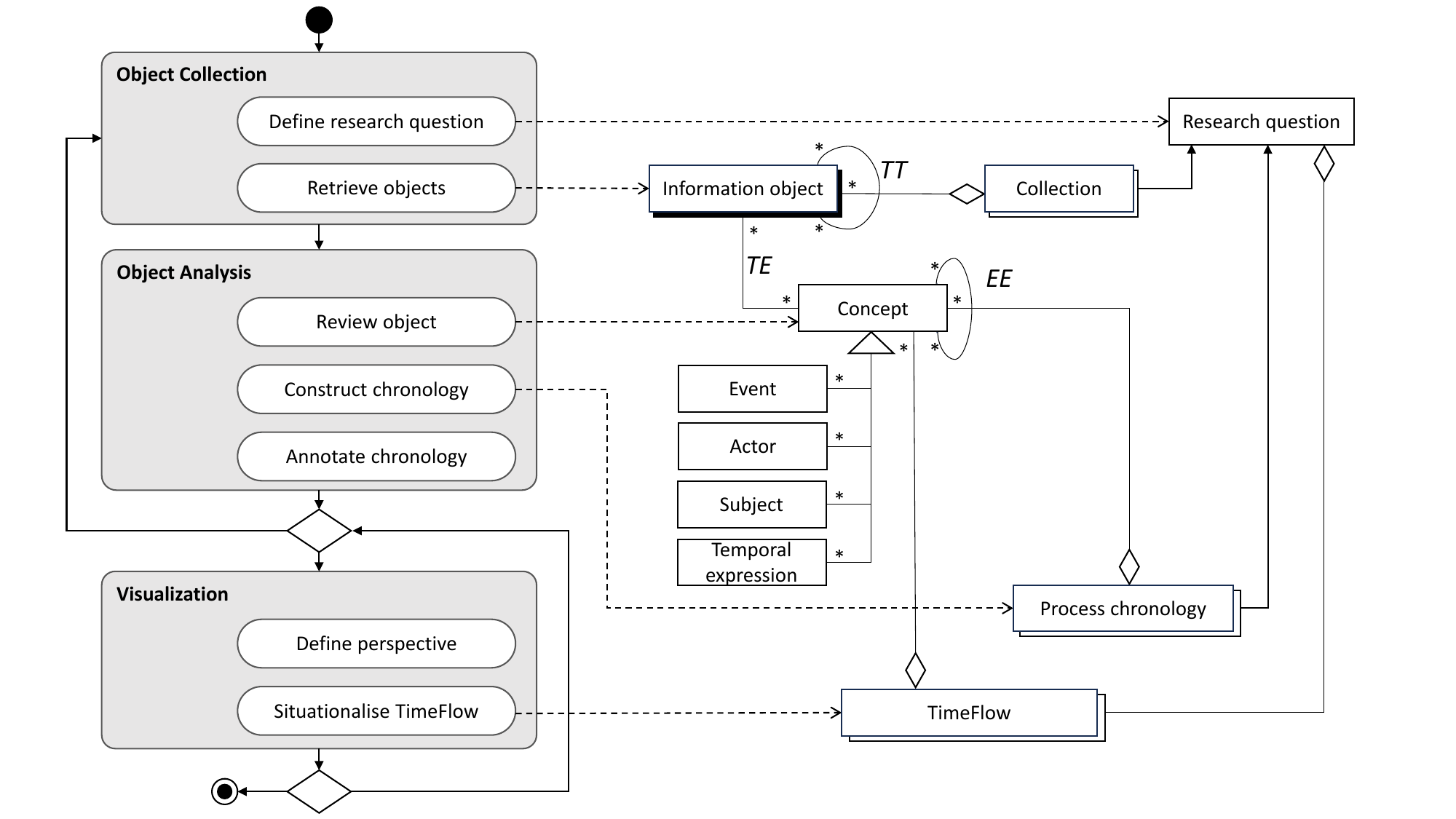}
\caption{Process Delivery Diagram of constructing TimeFlows based on the participants. The relations $\mathit{TT}$, $\mathit{TE}$ and $\mathit{EE}$ refer to the relations specified in Table~\ref{catgroup}. Together, these relations form a taxonomy. }\label{fig:pddoverview}
\end{figure}

During the interviews, we asked participants their current method of constructing chronologies. 
The answers of the participants have been coded to discover the different activities and sub activities. 
This resulted in the PDD shown in Fig.~\ref{fig:pddoverview}.
Each activity is supported by interviewee groups, indicated in parentheses. 

The first activity is \emph{Object Collection} (I1), which starts either by developing a research question for an information inquiry or by an external inquiry (I3).
For example, the parliament defines an inquiry, and then propagates the request to the different departments (I1, I4).
Typically, TimeFlows are requested to derive possible interpretations, for example to clarify the subjective images entities (including actors) can have in specific matters (I4).
Based on the research question, an initial collection of \textit{Information Objects} is retrieved from a plethora of sources, including Document Management Systems (I1) and open sources (I3).
An \textit{Information Object} can range from the agenda and minutes of meetings (I2), memos and email traffic (both within and between organisations) (I1), but can also include public sources such as news papers or social media (I3).
The \textit{Information Object} includes metadata about the object itself, such as creation data, version information and author information. 
The main challenge in this activity is that many information objects are unstructured, or poorly archived (I4).
For example, topics can be discussed in meetings, but to discover the records carrying this information can be painstaking (I1). 
Another problem is versioning, for example the detection and selection of near-duplicate information objects (I4). \par The next activity is \emph{Object Analysis}. In this activity, all retrieved \textit{Information Objects} are reviewed on relevance. This review is mostly a manual task ranging from scanning the object and its metadata to a careful read and interpretation (I1, I4). 
An initial scan is used to determine the relevance of each object (I1).
Relevant objects are then analysed to extract the most important \emph{concepts}, such as \emph{events}, \emph{entities}, \emph{subjects}, and \emph{temporal expressions} (I1, I3).

The extracted concepts are used to construct a \emph{chronology}: a large model containing all events extracted from the relevant objects.
Each event in the chronology is annotated with the relevant concepts, containing everything that could be relevant to it (I1), ranging from a single sentence of a document to complete documents (I4).
In rare cases, sub-chronologies are created to describe how specific documents were established (I2). 

The \emph{Object Analysis} activity is typically a manual task, and thus very laborious and time consuming (I1, I4).
As one of the participants stated: ``it is like looking for a needle in an exponentially growing haystack'' (I4).
Additionally, objects as policy documents require a thorough analysis to fully grasp the implications of the document (I3).
As such, the object analysis is subjective to the researchers who develop the chronology (I2).

The activities are very iterative, switching between the activities \emph{Object Collection} and \emph{Object Analysis}. 
Based on the chronology, new objects are retrieved to fill in gaps, to clarify events, or to provide an interpretation (I1).
For example, if in the agenda of a meeting specific documents are mentioned, these are retrieved and added to the collection for analysis (I2).

Currently, a \emph{Chronology} is described in a single document that contains all events and the concepts related to it. 
Such chronology documents easily grow to several hundreds of pages (I4).
As a consequence, it is infeasible to fully understand the events, and thus the process described in the document.
Instead, a specific perspective is created to construct an overview, for example to brief someone (I1, I2), or to gain a better understanding of the topic at hand (I4). 
For this activity, each event is manually screened and added, based on a set of criteria that is often left implicit and subject-dependent.
For example, if an event could not have been known by a stakeholder at the time, then the event is left out (I2).
Different perspectives result in situationalized representations of the chronology.
For example, a different representation is used for story telling (I2) than for understanding a certain topic (I4).
In the former situation, the representation mainly describes what happened, including important milestones and involved entities, whereas the latter may be a timeline to identify gaps, to refine specific concepts (I4).
As one of the participants noted: a chronology is never static (I4).

\mywidth{-4mm}
\subsection{Identified Relations Used in Document Analysis}\label{sec:foundrelations}

As became apparent from the interviews, \emph{Object Analysis} is the most laborious and time-consuming activity. 
All information objects need to be reviewed on relevance. Moreover, important concepts, such as events and entities, ought to be extracted.
This is a manual task, as their interpretation is far from trivial. It requires in-depth knowledge, which is often tacit (I3, I4).
As the participants indicated (I1, I3, I4), the concept extraction and chronology creation is a subjective task, where the researcher uses their tacit knowledge to decide the relevance of objects and the concepts it contains or discusses. 

Equally important are the relations that explain the link between information objects and concepts. 
We distinguish three levels of relations:
\begin{enumerate}
\item \textit{Between Information Objects}, i.e., relationships between different information objects (represented by level \emph{TT} in Fig.~\ref{fig:pddoverview}).
\item \textit{Between Information Objects and Concepts}, i.e., relationships that connect information objects to concepts (represented by level \emph{TE} in Fig.~\ref{fig:pddoverview}).
\item \textit{Between Concepts}, i.e., relationships between different concepts, such as events and entities (represented by level \emph{EE} in Fig.~\ref{fig:pddoverview}).
\end{enumerate}

\noindent
As shown in Section~\ref{literature}, different types of relations can be identified, such as information objects that describe a certain event, or entities that participated in a meeting described by the information object.
Therefore, we also asked the participants which relationships they use during the chronology construction.
We extracted eight relations that are commonly used by the participants to obtain concepts from information objects and relate them. 

\begin{figure}[t]
\centering
\includegraphics[width=0.65\textwidth]{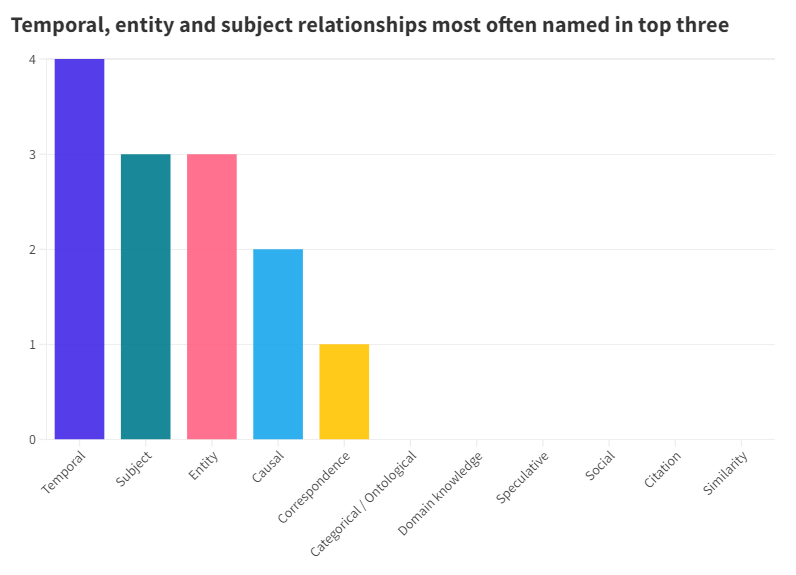} 
\caption{Number of interviewee groups that mention the respective relationships between events as one of the top three most important ones}
\label{top3}
\end{figure}

Each of the participants created a top 3 of relation types. The results are shown in Fig.~\ref{top3}: \emph{Temporal}, \emph{Subject}, and \emph{Entity} relationships were scored highest. The \emph{Causal} relationship and \emph{Correspondence} relations were mentioned less frequently. We choose to include all five of them within our framework. 

\begin{table}[b!]
\centering
\caption{Relations identified by the participants and their use over the different levels.}
\label{catgroup}
\begin{tabular}{|c |l |c| c| c|} 
 \hline
 &  & \  TT \ & \ EE \ & \ TE \ \\ [0.1ex] 
 \hline
 \multirow{5}{4em}{Group A} &  Temporal & x & x & x \\ 
 & Subject & x & x & x \\
 & Entity & x & x & x \\
 & Causal & x & x & x \\
 & Correspondence & x & x & x \\
 \hline
 \multirow{2}{4em}{Group B} & Succession & x & &  \\
 & References To & x &  &  \\ 
 \hline
 \multirow{1}{4em}{Group C} & Consists Of &  &  & x \\ [1ex] 
 \hline
\end{tabular}
\label{tablerelationships}
\end{table}

Our coding revealed three additional relations that were described by the participants, but not mentioned explicitly.
For example, all participants indicated they extracted the concepts from information objects.
In essence, this means that each information object is linked to several concepts that are extracted from it. 
We refer to this relation as the \emph{Consists Of} relation. This is the first of the three additional relations. 

Two other relations were also identified while coding the interviews.
In one of the interviews (I1), the participants explained that they considered email conversations, and added information based on the sender/receiver relation: ``If we look at the most important message [of a chronology], we also look at policy documents, but certainly at email conversations.'' His colleague added: ``Yes, so we look at who has responded to whom, and who has responded to that.''
Similarly, in another interview (I4), the participants indicated that they used the metadata of policy documents to derive the involved entities, and the dates on which these entities were involved.
This analysis resulted in two additional relations: \emph{Succession} and \emph{Reference To}.
The former indicates which objects directly follow each other, e.g. if an email is a direct reply to another email or if it is a forwarded message. The latter represents document referrals, such as documents discussed in meetings, or email messages with an attachment. 

We categorize all eight relations within three groups: A, B, and C, according to the types of relationships they afford (see Table \ref{tablerelationships}). In this upcoming section, we explain how these relations are incorporated within our conceptualization of process chronologies, and subsequently visualized within \emph{TimeFlows}.

\section{Visualizing TimeFlows} \label{framework}
\begin{figure}[t]
\centering
\includegraphics[width=0.7\textwidth]{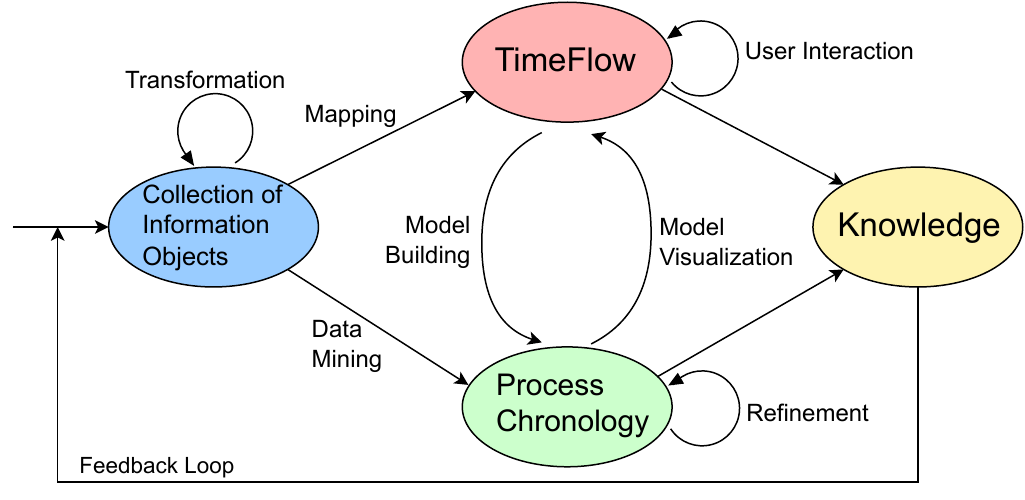}
\caption{The Visual Analytics feedback loop~\cite{visanalytics} applied to process chronologies. }
\label{vizanfeedback}
\end{figure}

In this section, we conceptualize process chronologies to construct TimeFlows from collections of information objects. 
As several participants in the interviews mentioned, visualising process chronologies has many different purposes, ranging from understanding the role of entities in the process (I2) to exploratory studies to identify gaps (I4).
The field of Visual Analytics ``combines automated analysis techniques with interactive visualizations for an effective understanding, reasoning and decision making on the basis of very large and complex data sets''~\cite{visanalytics}. 
Thus, visualizing TimeFlows can be considered as an application of Visual Analytics, as depicted in Fig.~\ref{vizanfeedback}. 
From the collection of information objects, a process chronology is extracted. 
This chronology is then visualized as a TimeFlow to gain new knowledge on the topic at hand.
In the remainder of this section, we present our proposal to visualize process chronologies as TimeFlows, and illustrate its capabilities by presenting a TimeFlow for the Dutch Childcare Benefits Scandal. 

\subsection{TimeFlows to Visualize Process Chronologies}

As presented in the previous section, a process chronology is a list of all annotated events extracted from a collection of information objects. Every event is annotated with the concepts it is related to, including entities and other events. Moreover, the events are enriched with the information objects that constitute them. A TimeFlow is a visual representation of a process chronology, to study a specific perspective. The following general design principles guide the visualization of TimeFlows: 

\begin{enumerate}[noitemsep]
    \item We delineate \emph{information objects} on the one hand, and \emph{events} on the other. In their simplest form, events can be seen as ``things that happen''. Events leave traces in the form of information objects that can be viewed as manifestations of past events \cite{ferejohn2013formal}. 
    Events are depicted using white rounded rectangles, while information objects are shown in light grey rectangles with sharp corners. 
    \item Relations are shown by means of arrows. The arrows stemming from the relations in group A are distinguished from one another through the use of both icons and colors. 
    The relations in group B and C are all shown in black, but have different stroke styles (solid, dashed, or dotted). 
    \item The progression of time is indicated by means of arrows from left to right between events, originating from relations in group A. Choosing to depict the flow of time from left to right stems from old conventions ingrained within Western culture \cite{timecontoursnunez}. Moreover, events are numbered according to their temporal order, which also makes it easy to refer to them. 
    \item The relationships depicted by the arrows in Group A are explained through the use of highlighted words in the information objects and events. They have the same color as the arrows of the respective relations, and indicate which words are of relevance to the relation. 
\end{enumerate}

\noindent The relations described in point 2 above are chosen and motivated in Section \ref{sec:foundrelations}. Below, we describe the design considerations for each individual relationship.

\begin{itemize}[noitemsep, topsep=2pt]
    \item \textbf{Temporal-Semantic Relationship}. \includegraphics[width=0.055\textwidth]{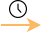} We relate events and information objects with one another that bear the same \emph{temporal expressions}. These are words that indicate a particular moment in time \cite{tempexprec2010}. 
    For instance, if one object contains the phrase ``On the \underline{first of September}, we had a meeting,'' and another object contains the fragment ``Things went well during the meeting on \underline{01/09}, we'll meet again soon,'' then these objects are linked through the Temporal-Semantic Relationship by means of a light orange arrow with a clock icon. Other examples of temporal expressions are ``tomorrow'', ``next week'', ``Halloween'', and ``The Summer holidays''. (Levels: TT, EE, and TE.)
    \item \textbf{Subject Relationship}.\includegraphics[width=0.055\textwidth] {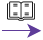} Events and information objects are linked with one another with a purple arrow and a book icon if the texts they comprise are related by subject. (Levels: TT, EE, and TE.) 
    \item \textbf{Entity Relationship}. \includegraphics[width=0.060\textwidth]{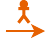} An Entity Relationship is made when the same concept, e.g. a name of a person, an organization, or a place, is mentioned in both of the events or objects. Those objects that belong to a particular event and mention the same concept, are assigned an Entity relationship by means of an orange arrow and a stick figure icon. (Levels: TT, EE, and TE.)
    \item \textbf{Causal Relationship}. \includegraphics[width=0.060\textwidth]{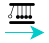} This relationship is visualized with a turquoise arrow and an icon of Newton's Cradle on top. It depicts which events are causally related to another, and is also drawn between information objects and events when words that indicate a causal relationship are present. (Levels: TT, EE, and TE.) 
    \item \textbf{Correspondence Relationship}. \includegraphics[width=0.060\textwidth]{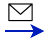} This relationship is depicted with a dark blue arrow and an icon of a letter on top. It is drawn when the correspondent of one object is also recognized as the correspondent of another object. (Levels: TT, EE, and TE.) 
    \item \textbf{Succession Relationship}. \includegraphics[width=0.055\textwidth]{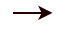} The \emph{Succession Relationship} is a black, filled arrow with its head pointing towards to an information object that is a response or forward of the original text. (Level: TT)
    \item \textbf{References To Relationship}. \includegraphics[width=0.055\textwidth]{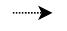} The \emph{References To Relationship} is depicted by means of a shortly-spaced dotted black arrow. It ends at the information object that is referenced, and begins at the object that does the referencing. It is employed when, for instance, an e-mail contains an attachment. (Level: TT)
    \item \textbf{Consists Of Relationship}. \includegraphics[width=0.055\textwidth]{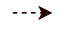} The \emph{Consists Of Relationship} is shown by means of a widely-spaced dotted black arrow, originating from the event and pointing towards the different information object the event consists of. (Level: TE)
\end{itemize} 

\noindent As intended by the Visual Analytics framework, it is the user that  constructs TimeFlows themselves, based on the information objects they have collected and their own perspective on the information objects they wish to study. 

\mywidth{-3mm}
\subsection{An Illustrative Example}
\begin{sidewaysfigure}
\includegraphics[width=0.8\textwidth, keepaspectratio]{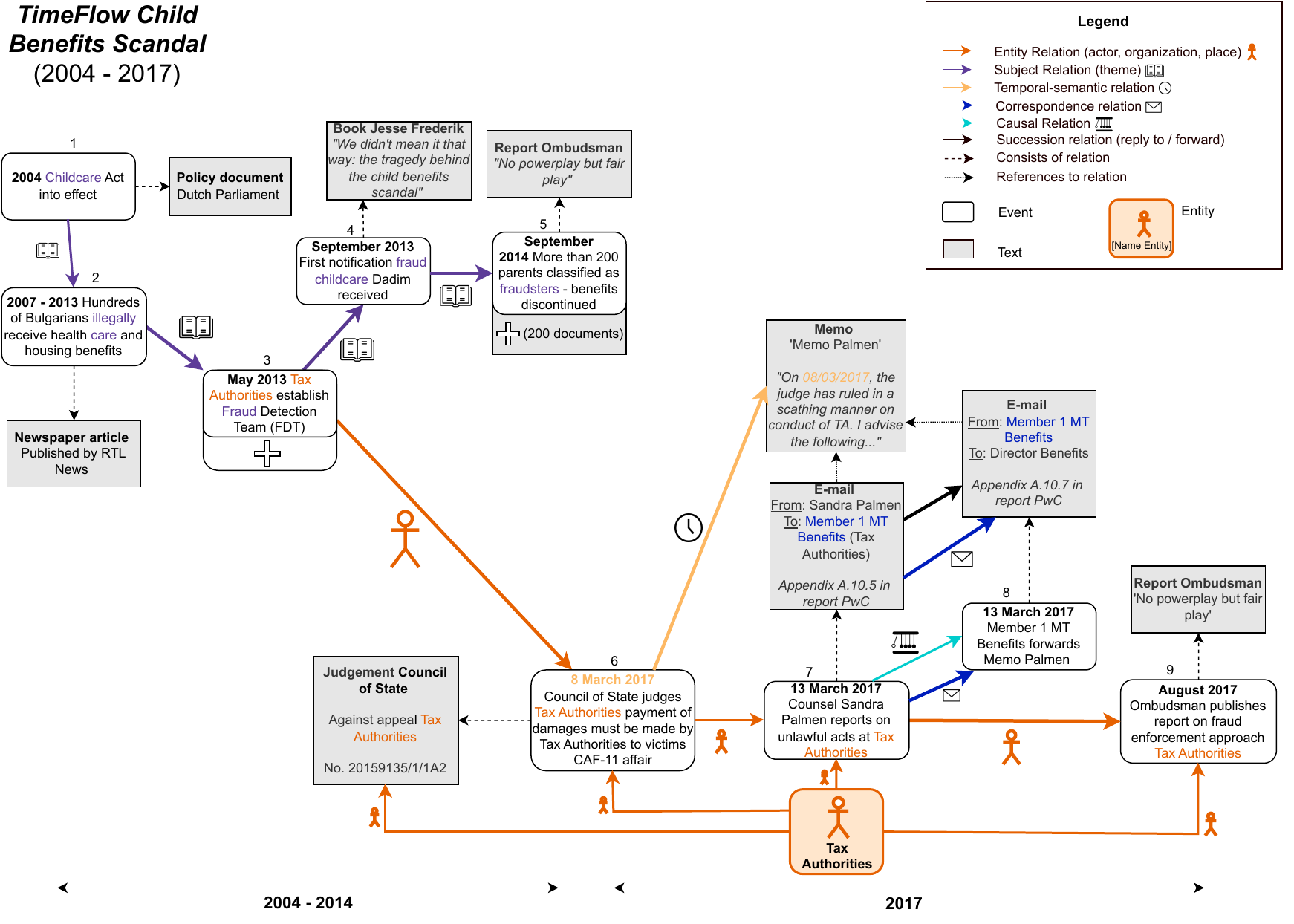}
\caption{A TimeFlow of the Childcare Benefits Scandal}
\end{sidewaysfigure} 

We illustrate the proposal of TimeFlows to represent process chronologies by depicting a sequence of events and accompanying documents pertaining to what is known in the Netherlands as the ``Childcare Benefits Scandal''. This political affair has its origins in the year 2004 (when the Childcare Act came into effect), and concerned false allegations of fraud made by the Tax Authorities. It is estimated that between the years 2005 and 2019, this organization wrongfully accused more than 25.000 parents of making fraudulent childcare benefits claims~\footnote{%
See \url{https://www.trouw.nl/politiek/wie-wist-wat-in-de-toeslagenaffaire-de-kluwen-van-hoofdrolspelers-ontward~b721c834/}%
}, driving families into severe financial hardship. After a parliamentary inquiry was conducted in 2021, the third cabinet Rutte resigned over the affair. 

This TimeFlow visualization is grounded in true events and real texts. A key document was the ``Memo Palmen'' \cite{rapportpwcbijlage2,rapportpwcbijlage1}. 
The other texts we used are, in order of appearance: the Dutch Childcare Act of 2004 \cite{childcare2004}, a news article on the `Bulgarian Fraud'~\footnote{See for example \url{https://www.rtlnieuws.nl/nieuws/nederland/artikel/1173981/uitspraak-bulgarenfraude-hoe-zat-het-ook-alweer}}, the book on the affair as a whole by Jesse Frederik \cite{frederik2021}, the report by the Ombudsman \cite{powerplay2017}, and the judgement of the Council of State \cite{uitspraakrvs}. We make the following general observations of this TimeFlow: 

\begin{itemize}[noitemsep, topsep=2pt] 
\item It consists of nine events. These are labelled by their temporal order. 
\item All events are distilled of one or more texts, shown with their respective Consists Of relations. In total, nine texts are explicitly shown. The existence of 200 additional texts is implied through the depiction of the + sign shown underneath Event 5.  
\item Note that Event 5 and Event 9 both consist of the report by the Ombudsman.  
\end{itemize}

\noindent
Moreover, additional information can be highlighted in such a TimeFlow. The current visualisation shows different examples of relations between events and information objects:

\begin{itemize}[noitemsep, topsep=2pt]
\item A Temporal-Semantic relation is indicated between Event 6 and the Memo Palmen, as they both refer to the date 08-03-2017 (TE). 
\item Subject relations are drawn between Event 1 and Event 2, Event 2 and Event 3, Event 3 and Event 4, and between Event 4 and Event 5 (EE). For instance, the Subject relation between the first two events is justified because they contain the related words ``Childcare'' and ``care''. 
\item Entity relations are depicted between Event 3 and Event 6, Event 6 and Event 7, and between Event 7 and Event 9 (EE). We also single out the Tax Authorities as an entity concept, and include relations to the judgement of the Council of State (TE) and to Events 6, 7, and 9 (EE). 
\item A Causal relation links events Events 7 and 8, because the act of forwarding the email - performed by Member 1 of the MT - is a direct causal consequence of Event 7, in which Sarah Palmen shares her findings (EE).
\item Events 7 and 8 are also connected via the Correspondence relation, because the latter consists of an email that is forwarded from an email Event 7 consists of (EE).  
\item A Succession relation is shown between the Sarah Palmen's email, and the forwarded email sent by Member 1 of the MT of the Tax Authorities (TT). 
\item Both Palmen's and MT Member 1's emails contain Palmen's report as an attachment - thus both of these emails point to the report with the References To relation (TT). 
\end{itemize}

\noindent As the events in this example show, multiple relation types are needed to understand and study what happened. As shown in Section~\ref{literature}, these relations are not shown in current visualisation techniques, such as the TGR in Figure~\ref{storytree2016}. 
Moreover, the constitutive documents clarify on what basis the events have been generated -- something that is lacking in previous conceptualizations of timelines with graph-based representations (TGRs).

\section{Further Research: Challenges} \label{challenges}
In this paper, we argue that the visual analytics framework (see Figure \ref{vizanfeedback}) can be applied to study process chronologies using TimeFlows. 
However, to realize this, many challenges still need to be addressed. These challenges originate from the problems described by the interviewees.   \noindent

\begin{description}[noitemsep, topsep=2pt, leftmargin=0cm
]
    \item[Data Mining] The future software ought to be able to automatically mine important concepts of heterogeneous document sets. Relevant challenges in this context include \emph{Named Entity Recognition and Linking} (NER and NEL), \emph{Temporal Expression Recognition, Normalization, and Linking} (TERNL), \emph{Topic Modelling} (TM), and \emph{Event Extraction} (EvEx).  
    \item[Mapping] The relations between the properties of the input data (information objects) and the visualizations (TimeFlows) are to be studied. A challenge lies in mapping more heterogeneous data sources to proper visualizations, including tabular data in Excel sheets, meeting minutes, and voice memos. 
    \item[Model Visualization] In order that the TimeFlows are generated in a visually intuitive manner, research must be carried out on peoples' preferred shapes and colours, the positioning of the events, information objections, and relations, and the amount of relations shown so as to balance clarity on the one hand, yet avoid information overload on the other.
    \item[User Interaction] Here, the challenge lies in understanding how users can be supported in developing effective TimeFlows, e.g., the interaction with TimeFlows to enable users to obtain the information objects that bear relevance to each event, to move concepts and entities on the screen, to `to focus on' or `abstract away' specific events and investigate them at different levels of granularity.  
    \item[Automated Model Building] Future TimeFlows software could allow users to study different perspectives on sequences of events and their information objects. They may single out a perspective or combine different perspectives to study a research question from different angles. To support this activity, generative techniques need to be developed that propose multiple perspectives on the collection of information objects. These perspectives 
    can help in the understanding of the process and the events manifested within the information objects.  
\end{description}

\section{Conclusion} \label{conclusion}
In this paper, we introduce \emph{process chronologies} as non-repetitive processes, which lack an event log. 
Based on interviews with experts, we have conceptualized different relations between the concepts and unstructured information objects that underlie them. 
To convey the events in a non-repetitive process, we propose \emph{TimeFlows} to visualize the relation between information objects and the concepts manifested in these objects. 
TimeFlows allow users to study non-repetitive processes from different perspectives.
Although our analysis was conducted on explorative interviews with domain experts, we believe that an holistic approach that visual analytics with TimeFlows offers great potential aid the understanding of such processes. 
However, to fully support the understanding and study of non-repetitive processes, many challenges need to be overcome. We therefore propose a research agenda to guide future work in this direction, which is aimed at automating the time-consuming manual work for process chronology creation.
We view the present work as a broader ambition to develop interactive software that allows analysts to inspect large amounts of heterogeneous data sources to retrospectively study how non-repetitive processes unfold over time.

%
%
\bibliographystyle{ACM-Reference-Format}
\bibliography{bibliography}


\begin{thebibliography}{21}


\ifx \showCODEN    \undefined \def \showCODEN     #1{\unskip}     \fi
\ifx \showDOI      \undefined \def \showDOI       #1{#1}\fi
\ifx \showISBNx    \undefined \def \showISBNx     #1{\unskip}     \fi
\ifx \showISBNxiii \undefined \def \showISBNxiii  #1{\unskip}     \fi
\ifx \showISSN     \undefined \def \showISSN      #1{\unskip}     \fi
\ifx \showLCCN     \undefined \def \showLCCN      #1{\unskip}     \fi
\ifx \shownote     \undefined \def \shownote      #1{#1}          \fi
\ifx \showarticletitle \undefined \def \showarticletitle #1{#1}   \fi
\ifx \showURL      \undefined \def \showURL       {\relax}        \fi
\providecommand\bibfield[2]{#2}
\providecommand\bibinfo[2]{#2}
\providecommand\natexlab[1]{#1}
\providecommand\showeprint[2][]{arXiv:#2}

\bibitem[Ansah et~al\mbox{.}(2019)]%
        {ansah2019}
\bibfield{author}{\bibinfo{person}{Jeffery Ansah}, \bibinfo{person}{Lin Liu},
  \bibinfo{person}{Wei Kang}, \bibinfo{person}{Selasie Kwashie},
  \bibinfo{person}{Jixue Li}, {and} \bibinfo{person}{Jiuyong Li}.}
  \bibinfo{year}{2019}\natexlab{}.
\newblock \showarticletitle{A graph is worth a thousand words: Telling event
  stories using timeline summarization graphs}. In
  \bibinfo{booktitle}{\emph{The World Wide Web Conference}}.
  \bibinfo{pages}{2565--2571}.
\newblock


\bibitem[{Dutch Government}(2004)]%
        {childcare2004}
\bibfield{author}{\bibinfo{person}{{Dutch Government}}.}
  \bibinfo{year}{2004}\natexlab{}.
\newblock \showarticletitle{Wet Kinderopvang}.
\newblock \bibinfo{journal}{\emph{Wettenbank}} (\bibinfo{date}{Oct.}
  \bibinfo{year}{2004}).
\newblock


\bibitem[Ferejohn(2013)]%
        {ferejohn2013formal}
\bibfield{author}{\bibinfo{person}{Michael~T Ferejohn}.}
  \bibinfo{year}{2013}\natexlab{}.
\newblock \bibinfo{booktitle}{\emph{Formal Causes: Definition, Explanation, and
  Primacy in Socratic and Aristotelian Thought}}.
\newblock \bibinfo{publisher}{OUP Oxford}.
\newblock


\bibitem[Frederik(2021)]%
        {frederik2021}
\bibfield{author}{\bibinfo{person}{J. Frederik}.}
  \bibinfo{year}{2021}\natexlab{}.
\newblock \bibinfo{booktitle}{\emph{Zo hadden we het niet bedoeld: De tragedie
  achter de toeslagenaffaire}}.
\newblock \bibinfo{publisher}{De Correspondent Uitgevers}. 1--400 pages.
\newblock


\bibitem[Keim et~al\mbox{.}(2008)]%
        {visanalytics}
\bibfield{author}{\bibinfo{person}{D. Keim}, \bibinfo{person}{G. Andrienko},
  \bibinfo{person}{J.-D. Fekete}, \bibinfo{person}{C. Görg},
  \bibinfo{person}{J. Kohlhammer}, {and} \bibinfo{person}{G. Melancon}.}
  \bibinfo{year}{2008}\natexlab{}.
\newblock \showarticletitle{Visual Analytics: Definition, Process, and
  Challenges}.
\newblock \bibinfo{journal}{\emph{Information Visualization}}
  \bibinfo{volume}{LNCS}, \bibinfo{number}{4950} (\bibinfo{date}{Sept.}
  \bibinfo{year}{2008}), \bibinfo{pages}{154--175}.
\newblock


\bibitem[Keith~Norambuena and Mitra(2021)]%
        {noram2021}
\bibfield{author}{\bibinfo{person}{Brian~Felipe Keith~Norambuena} {and}
  \bibinfo{person}{Tanushree Mitra}.} \bibinfo{year}{2021}\natexlab{}.
\newblock \showarticletitle{Narrative Maps: An Algorithmic Approach to
  Represent and Extract Information Narratives}.
\newblock \bibinfo{journal}{\emph{Proc. ACM Hum.-Comput. Interact.}}
  \bibinfo{volume}{4}, \bibinfo{number}{CSCW3}, Article
  \bibinfo{articleno}{228} (\bibinfo{date}{jan} \bibinfo{year}{2021}),
  \bibinfo{numpages}{33}~pages.
\newblock


\bibitem[Keith~Norambuena et~al\mbox{.}(2022)]%
        {noram2022}
\bibfield{author}{\bibinfo{person}{B.~F. Keith~Norambuena}, \bibinfo{person}{T.
  Mitra}, {and} \bibinfo{person}{C. North}.} \bibinfo{year}{2022}\natexlab{}.
\newblock \showarticletitle{Design guidelines for narrative maps in sensemaking
  tasks}.
\newblock \bibinfo{journal}{\emph{Information Visualization}}
  \bibinfo{volume}{21}, \bibinfo{number}{3} (\bibinfo{date}{March}
  \bibinfo{year}{2022}), \bibinfo{pages}{220--245}.
\newblock


\bibitem[Keith~Norambuena et~al\mbox{.}(2023)]%
        {noram2023}
\bibfield{author}{\bibinfo{person}{B.~F. Keith~Norambuena}, \bibinfo{person}{T.
  Mitra}, {and} \bibinfo{person}{C. North}.} \bibinfo{year}{2023}\natexlab{}.
\newblock \showarticletitle{A Survey on Event-based News Narrative Extraction}.
\newblock \bibinfo{journal}{\emph{Comput. Surveys}} \bibinfo{volume}{55},
  \bibinfo{number}{14s} (\bibinfo{date}{July} \bibinfo{year}{2023}),
  \bibinfo{pages}{1--39}.
\newblock


\bibitem[Liu et~al\mbox{.}(2020)]%
        {liu2020}
\bibfield{author}{\bibinfo{person}{B. Liu}, \bibinfo{person}{F.~X. Han},
  \bibinfo{person}{D. Niu}, \bibinfo{person}{L. Kong}, \bibinfo{person}{K.
  Lai}, {and} \bibinfo{person}{Y. Xu}.} \bibinfo{year}{2020}\natexlab{}.
\newblock \showarticletitle{Story Forest: Extracting Events and Telling Stories
  from Breaking News}.
\newblock \bibinfo{journal}{\emph{ACM Transactions on Knowledge Discovery from
  Data}} \bibinfo{volume}{14}, \bibinfo{number}{3} (\bibinfo{date}{May}
  \bibinfo{year}{2020}), \bibinfo{pages}{1--28}.
\newblock


\bibitem[Núñez et~al\mbox{.}(2012)]%
        {timecontoursnunez}
\bibfield{author}{\bibinfo{person}{R. Núñez}, \bibinfo{person}{K.
  Cooperrider}, \bibinfo{person}{D. Doan}, {and} \bibinfo{person}{J.
  Wassmann}.} \bibinfo{year}{2012}\natexlab{}.
\newblock \showarticletitle{Contours of time: Topographic construals of past,
  present, and future in the Yupno valley of Papua New Guinea}.
\newblock \bibinfo{journal}{\emph{Cognition}} \bibinfo{volume}{124},
  \bibinfo{number}{1} (\bibinfo{date}{July} \bibinfo{year}{2012}),
  \bibinfo{pages}{25--35}.
\newblock


\bibitem[PricewaterhouseCoopers(2021a)]%
        {rapportpwcbijlage2}
\bibfield{author}{\bibinfo{person}{PricewaterhouseCoopers}.}
  \bibinfo{year}{2021}\natexlab{a}.
\newblock \showarticletitle{Appendices bij PwC rapport Reconstructie en
  tijdlijn van het ‘memo-Palmen’}.
\newblock \bibinfo{journal}{\emph{Report PwC}} (\bibinfo{date}{Sept.}
  \bibinfo{year}{2021}), \bibinfo{pages}{1--301}.
\newblock


\bibitem[PricewaterhouseCoopers(2021b)]%
        {rapportpwcbijlage1}
\bibfield{author}{\bibinfo{person}{PricewaterhouseCoopers}.}
  \bibinfo{year}{2021}\natexlab{b}.
\newblock \showarticletitle{Reconstructie en tijdlijn van het
  ‘memo-Palmen’}.
\newblock \bibinfo{journal}{\emph{Report PwC}} (\bibinfo{date}{Sept.}
  \bibinfo{year}{2021}), \bibinfo{pages}{1--66}.
\newblock


\bibitem[{Raad van State (Council of State)}(2017)]%
        {uitspraakrvs}
\bibfield{author}{\bibinfo{person}{{Raad van State (Council of State)}}.}
  \bibinfo{year}{2017}\natexlab{}.
\newblock \showarticletitle{Uitspraak 201509135/1/A2}.
\newblock \bibinfo{journal}{\emph{https://www.raadvanstate.nl/}}
  (\bibinfo{year}{2017}), \bibinfo{pages}{1--8}.
\newblock


\bibitem[Shahaf and Guestrin(2011)]%
        {shahaf2010}
\bibfield{author}{\bibinfo{person}{Dafna Shahaf} {and} \bibinfo{person}{Carlos
  Guestrin}.} \bibinfo{year}{2011}\natexlab{}.
\newblock \showarticletitle{Connecting the Dots between News Articles}. In
  \bibinfo{booktitle}{\emph{{IJCAI} 2011}}. \bibinfo{publisher}{{IJCAI/AAAI}},
  \bibinfo{pages}{2734--2739}.
\newblock


\bibitem[Shahaf et~al\mbox{.}(2012)]%
        {shahaffinfomaps}
\bibfield{author}{\bibinfo{person}{Dafna Shahaf}, \bibinfo{person}{Carlos
  Guestrin}, {and} \bibinfo{person}{Eric Horvitz}.}
  \bibinfo{year}{2012}\natexlab{}.
\newblock \showarticletitle{Trains of thought: generating information maps}. In
  \bibinfo{booktitle}{\emph{{WWW} 2012}}. \bibinfo{publisher}{{ACM}},
  \bibinfo{pages}{899--908}.
\newblock


\bibitem[Steen and Markert(2019)]%
        {steen2019}
\bibfield{author}{\bibinfo{person}{Julius Steen} {and} \bibinfo{person}{Katja
  Markert}.} \bibinfo{year}{2019}\natexlab{}.
\newblock \showarticletitle{Abstractive Timeline Summarization}. In
  \bibinfo{booktitle}{\emph{Proceedings of the 2nd Workshop on New Frontiers in
  Summarization}}. \bibinfo{publisher}{Association for Computational
  Linguistics}, \bibinfo{address}{Hong Kong, China}, \bibinfo{pages}{21--31}.
\newblock


\bibitem[UzZaman and Allen(2010)]%
        {tempexprec2010}
\bibfield{author}{\bibinfo{person}{N. UzZaman} {and} \bibinfo{person}{J.~F.
  Allen}.} \bibinfo{year}{2010}\natexlab{}.
\newblock \showarticletitle{Extracting Events and Temporal Expressions from
  Text}.
\newblock \bibinfo{journal}{\emph{2010 IEEE ICSC}} (\bibinfo{date}{Sept.}
  \bibinfo{year}{2010}), \bibinfo{pages}{1--8}.
\newblock


\bibitem[van~de Weerd et~al\mbox{.}(2006)]%
        {vandeweerd_2006_PDD}
\bibfield{author}{\bibinfo{person}{Inge van~de Weerd}, \bibinfo{person}{Sjaak
  Brinkkemper}, \bibinfo{person}{Jurriaan Souer}, {and} \bibinfo{person}{Johan
  Versendaal}.} \bibinfo{year}{2006}\natexlab{}.
\newblock \showarticletitle{A situational implementation method for web-based
  content management system-applications: method engineering and validation in
  practice}.
\newblock \bibinfo{journal}{\emph{Softw. Process. Improv. Pract.}}
  \bibinfo{volume}{11}, \bibinfo{number}{5} (\bibinfo{year}{2006}),
  \bibinfo{pages}{521--538}.
\newblock


\bibitem[van~den Berg et~al\mbox{.}(2017)]%
        {powerplay2017}
\bibfield{author}{\bibinfo{person}{W. van~den Berg}, \bibinfo{person}{M.
  Alhadjri}, {and} \bibinfo{person}{M. Mulder}.}
  \bibinfo{year}{2017}\natexlab{}.
\newblock \showarticletitle{Geen Powerplay, maar Fair Play}.
\newblock \bibinfo{journal}{\emph{De Nationale Ombudsman}}
  (\bibinfo{date}{Aug.} \bibinfo{year}{2017}).
\newblock


\bibitem[Xu et~al\mbox{.}(2013)]%
        {xu2013}
\bibfield{author}{\bibinfo{person}{Shize Xu}, \bibinfo{person}{Shanshan Wang},
  {and} \bibinfo{person}{Yan Zhang}.} \bibinfo{year}{2013}\natexlab{}.
\newblock \showarticletitle{Summarizing Complex Events: a Cross-Modal Solution
  of Storylines Extraction and Reconstruction}. In
  \bibinfo{booktitle}{\emph{EMNLP}}. \bibinfo{publisher}{Association for
  Computational Linguistics}, \bibinfo{address}{Seattle, Washington, USA},
  \bibinfo{pages}{1281--1291}.
\newblock


\bibitem[Yu et~al\mbox{.}(2021)]%
        {yu2021}
\bibfield{author}{\bibinfo{person}{Yi Yu}, \bibinfo{person}{Adam Jatowt},
  \bibinfo{person}{Antoine Doucet}, \bibinfo{person}{Kazunari Sugiyama}, {and}
  \bibinfo{person}{Masatoshi Yoshikawa}.} \bibinfo{year}{2021}\natexlab{}.
\newblock \showarticletitle{Multi-{T}ime{L}ine Summarization ({MTLS}):
  Improving Timeline Summarization by Generating Multiple Summaries}. In
  \bibinfo{booktitle}{\emph{IJCNLP}}. \bibinfo{publisher}{Association for
  Computational Linguistics}, \bibinfo{address}{Online},
  \bibinfo{pages}{377--387}.
\newblock


\end{thebibliography}






\end{document}